# CEPC Input to the ESPP 2018

# -Accelerator

## CEPC Accelerator Study Group

**Executive summary**

The discovery of the Higgs boson at CERN's Large Hadron Collider (LHC) in July 2012 raised new opportunities for a large-scale accelerator. Due to the low mass of the Higgs, it is possible to produce it in the relatively clean environment of a circular electron–positron collider with reasonable luminosity, technology, cost and power consumption. The Higgs boson is a crucial cornerstone of the Standard Model (SM). It is at the center of some of its biggest mysteries, such as the large hierarchy between the weak scale and the Planck scale, the nature of the electroweak phase transition, and many other related questions. Precise measurements of the properties of the Higgs boson serve as excellent tests of the underlying fundamental physics principles of the SM, and they are instrumental in explorations beyond the SM. In September 2012, Chinese scientists proposed a 240 GeV *Circular Electron Positron Collider* (CEPC), serving two large detectors for Higgs studies. The tunnel for such a machine could also host a *Super Proton Proton Collider* (SPPC) to reach energies beyond the LHC.

The CEPC is a large international scientific project initiated and hosted by China. It was presented for the first time to the international community at the ICFA Workshop "*Accelerators for a Higgs Factory: Linear vs. Circular*" (HF2012) in November 2012 at Fermilab. A Preliminary Conceptual Design Report (Pre-CDR, the *White Report*)[1]was published in March 2015, followed by a Progress Report (the *Yellow Report*)[2] in April 2017, where CEPC accelerator baseline choice was made. The Conceptual Design Report (CEPC Accelerator CDR, the *Blue Report*) [3]has been completed in July 2018 by hundreds of scientists and engineers after international review from June 28-30, 2018 and formally released on Sept 2, 2018.

The CEPC is a circular *e+e-* collider located in a 100-km circumference tunnel beneath the ground. The accelerator complex consists of a linear accelerator (Linac), a damping ring (DR), the Booster, the Collider and several transfer lines. In the tunnel, space is reserved for a future *pp* collider, SPPC. The center-of-mass energy of the CEPC is set at 240 GeV, and at that collision energy CEPC will serve as a Higgs factory, generating more than one million Higgs particles. The design also allows for operation at 91 GeV as a Z factory and at 160 GeV as a W factory. The number of Z particles produced will be close to one trillion, and W+W-pairs close to 20 million. The heart of the CEPC is a double-ring collider (except at SCRF region, where electron and positron use common beam pipe). Electron and positron beams circulate in opposite directions in separate beam pipes but with the common SCRF system. They collide at two interaction points (IPs), where large detectors as described in detail in the CDR (Volume II) are located. The CEPC Booster is located in the same tunnel above the Collider. It is a synchrotron with a 10 GeV injection energy and extraction energy equal to the beam collision energy. The repetition cycle is 10 seconds. Top-up injection will be used to maintain constant luminosity. The 10 GeV Linac, injector to the Booster, built at ground level, accelerates both electrons and positrons. A 1.1 GeV damping ring reduces the positron emittance. Transport lines made of permanent magnets connect the Linac to the Booster. The tunnel size is large enough to accommodate the future SPPC without removing the CEPC collider ring. This opens up the exciting possibilities of *ep* and *e*-ion physics in addition to *ee* physics (CEPC) and *pp* and ion-ion physics (SPPC). In addition to particle physics, the Collider can operate simultaneously as a powerful synchrotron radiation (SR) light source. It will extend the usable SR spectrum into an unprecedented energy and brightness range. Two gamma-ray beamlines are included in the design. The circulating CEPC beams radiate large amount of SR power, 30 MW per beam. Reducing power consumption is an important criterion in the design. By using superconducting radio frequency (SCRF) cavities, high efficiency klystrons, 2-in-1 magnets, combined function magnets, large coil cross-section in the quadrupoles, the total facility power consumption is kept below 300 MW. The power conversion efficiency from the grid to the beam will be more than 20%, higher than other accelerator facilities. Prior to the construction will be a five-year R&D period (2018-2022). During this period, prototypes of key technical components will be built and infrastructure established for industrialization for manufacturing the large number of required components. There are numerous considerations in choosing the site. At this moment six sites have been considered and they all satisfy the technical requirements. A detailed cost estimate based on a Work Breakdown Structure (WBS) has been carried out. CEPC Construction is expected to start in 2022 and be completed in 2030. After commissioning, a tentative operation plan will be running 7 years for Higgs physics, followed by 2 years for operation in Z mode and 1 year for operation in W mode. The large number of particles produced makes the CEPC a powerful instrument not only for precision measurements on these important particles, but also in the search for new physics. The CEPC is an important part of the world plan for high-energy physics research. It will support a comprehensive research program by scientists from throughout the world. Physicists from many countries will work together to explore the science and technology frontier, and to bring a new level of our understanding of the fundamental nature of matter, energy and the Universe.

**Parameters and the optimization design**

According to the CEPC physics goals at the Higgs and Z-pole energies, the CEPC should provide e+e- collisions at the center-of-mass energy of 240 GeV and deliver a peak luminosity of $2\times10^{34}$ cm$^{-2}$s$^{-1}$ at each interaction point. The CEPC has two IPs for e+e- collisions. At the Z-pole the luminosity is required to be larger than $1\times10^{34}$ cm$^{-2}$s$^{-1}$ per IP. Its circumference is around 100 km in accordance with the SppC, which is designed to provide proton-proton collisions at 100 TeV center-of-mass energy using 16 Tesla superconducting dipole magnets.

The CEPC baseline design is a 100 km fully partial double ring scheme based on crab waist collision and 30 MW radiation power per beam at Higgs energy, with the shared RF system for both electron and positron beams. As an alternative option, Advanced Partial Double Ring (APDR) has been also studied systematically with the aim of comparing the luminosity potentials and saving cost.

The luminosities for Higgs and W operation are mainly limited by the SR power (30 MW). The luminosity at Higgs is $3 \times 10^{34}$ cm$^{-2}$s$^{-1}$ with 242 bunches; the luminosity at the W is $1 \times 10^{35}$ cm$^{-2}$s$^{-1}$ with 1524 bunches. At the Z pole, the luminosity for 3T detector solenoid is $1.7 \times 10^{35}$ cm$^{-2}$s$^{-1}$ and is $3.2 \times 10^{35}$ cm$^{-2}$s$^{-1}$ for 2T detector solenoid, both with 12,000 bunches. The limit of bunch number comes from the electron cloud instability of the positron beam. The minimum bunch separation for Z due to electron cloud effect is 25 ns and a 10% beam gap is left for cleaning. There is still some space for parameter optimization at z pole to get even higher luminosity. Beam-beam interaction in is one of the most important limitations to the CEPC performance, which is calculated both by analytical method and computer simulations to make optimized parameter design and choose optimum operating conditions. The crab-waist scheme increases the luminosity by suppressing vertical blow up, which is a must to reach high luminosity. Beamstrahlung is synchrotron radiation excited by the beam-beam force, which is a new phenomenon in a storage ring based collider. It will increase the energy spread, lengthen the bunch and may reduce the beam lifetime due to the long tail of the photon spectrum. The beam-beam limit at the W/Z is mainly determined by the coherent x-z instability instead of the beamstrahlung lifetime as in the Higgs mode. Longer bunch length will help to suppress the coherent instability.

The CEPC CDR design goals have been evaluated and checked from the point view of beam-beam interaction, which is feasible and achievable.

**Lattice optics**

The CEPC lattice optics is designed with requirements and constraints mainly from top-level parameters, geometry, minimizing cost, compatibility of Higgs, W and Z modes, and compatibility with SPPC. The interaction region is designed to provide strong focusing and crab-waist collision. A local correction scheme is adopted to get a large momentum acceptance. An asymmetric lattice is adopted to allow softer synchrotron-radiation photons from the upstream part of the IP. Twin-aperture dipoles and quadrupoles are used in the arc region to reduce power. The two beams are separated by 35 cm. For the arc region, a FODO cell structure is chosen to provide a large filling factor of dipoles. The 90/90-degrees phase advances and non-interleaved-sextupole scheme are selected due to aberration cancellation. In the RF region, the RF cavities are shared by the two rings. Each RF station is divided into two sections for bypassing half of the cavities when running in W or Z modes. An electrostatic separator combined with a dipole magnet avoids bending the incoming beam. The sawtooth effect is expected to be curable by tapering the magnet strength to take into account the beam energy at each magnet. The vertical emittance due to the solenoid field coupling is limited and acceptable.

The requirements of dynamic aperture (DA) are got from injection and beam-beam effects to get efficient injection and adequate beam life time. A differential evolution algorithm based optimization code has been developed for CEPC, which is a multi-objective code called MODE. The SAD code is used to do the optics calculation and dynamic aperture tracking. Strong synchrotron radiation causes strong radiation damping which helps enlarge the dynamic aperture to some extent. Quantum fluctuations in the synchrotron radiation are considered in SAD, where the random diffusion due to synchrotron radiation in the particle tracking is implemented in each magnet. Thirty-two arc sextupole families, 10 IR sextupole families and 8 phase advance tuning knobs between different sections are used to optimize the DA. All the sextupoles (~250) could be free. There exists no clear difference with more sextupole families. The optimized DA could meet the requirement of injection and colliding beam lifetime. The work of DA with errors and correction are undergoing.

**Machine-detector interface (MDI)**

The machine-detector interface (MDI) is about ±7 m in length in the Interaction Region (IR),

where many elements need to be installed, including the detector solenoid, anti-solenoide, luminosity calorimeter, interaction region beam pipe, beryllium pipe, cryostat, beam position monitors (BPMs) and bellows. The cryostat includes the final doublet superconducting magnets and anti-solenoid. The CEPC detector consists of a cylindrical drift chamber surrounded by an electromagnetic calorimeter, which is immersed in a 3T superconducting solenoid of length 7.6 m. The accelerator components inside the detector should not interfere with the devices of the detector. The smaller the conical space occupied by accelerator components, the better will be the geometrical acceptance of the detector. From the requirement of detector, the conical space with an opening angle should not larger than 8.11 degrees. After optimization, the accelerator components inside the detector without shielding are within a conical space with an opening angle of 6.78 degrees. The crossing angle between electron and positron beams is 33 mrad in horizontal plane. The final focusing quadrupole is 2.2 m ($L^*$) from the Interaction Point (IP). The luminosity calorimeter will be installed in a longitudinal location 0.95~1.11 m, with an inner radius of 28.5 mm and outer radius 100 mm. Primary results are got from the assembly, interfaces with the detector hardware, cooling channels, vibration control of the cryostats, supports and so on.

A water cooling structure is required to control the heating problem of HOM in IR vacuum chamber. For the beam pipe within the final doublet quadrupoles, since there is a 4mm gap between the outer space of beam pipe and the inner space of Helium vessel, a room temperature beam pipe has been chosen.

SR photons in the IR are mainly generated from the final upstream bending magnet and the IR quadrupole magnets due to eccentric particles. With 3 mask tips along the inside of the beam pipe to shadow the inner surface of the pipe the number of scattered photons that can hit the central beam pipe is greatly reduced to only those photons which forward scatter through the mask tips. With collimators in the ARC far from IP the SR photons from IR quadrupoles will not damage the detector components and cause background to experiments.

Beam loss background are mainly from Bhabha scattering, beamstrahlung, beam-thermal photon scattering and beam-gas inelastic scattering. With collimators in the ARC far from IP, beam loss background can be reduced significantly and can be accepted by the detector.

**CEPC SCRF system**

CEPC will use a 650 MHz RF system with 240 2-cell superconducting cavities for the Collider and a 1.3 GHz RF system with 96 9-cell superconducting cavities for the Booster. The Collider is a fully partial double-ring with common cavities for electron and positron beams in Higgs operation mode and a double ring for separate cavities for electron and positron beams in W and Z operation modes. Full installation of the same types of cavities and cryomodules for Higgs, W, and Z-pole modes without changing any hardware is the baseline configuration. The Collider SCRF system is optimized for the Higgs mode of 30 MW SR power per beam, with enough tunnel space and operating margin to allow higher RF voltage and/or SR power (50 MW SR power per beam) by adding cavities. Each Collider cavity has two detachable coaxial HOM couplers mounted on the cavity beam pipe with HOM power handling capacity of 1 kW. Each 11 m-long cryomodule consists of six cavities. Each cryomodule has two beamline HOM absorbers at room temperature outside the vacuum vessel with HOM power handling capacity of 5 kW each.

HOM power limit per cavity and the fast-growing longitudinal coupled-bunch instabilities (CBI) driven by both the fundamental and higher order modes impedance of the RF cavities determine to a large extent the highest beam current and luminosity obtainable in the Z mode. Transient beam loading is also a concern. For a higher luminosity Z upgrade, because of the high HOM power and the need to have the smallest number of cavities, KEKB / BEPCII type single cavity cryomodules with very high input coupler power will be needed.

The CEPC SCRF technical challenges that require R&D include: achieving the cavity gradient and high quality factor in the real cryomodule environment ($Q_0 > 4 \times 10^{10}$ at 22 MV/m for the vertical acceptance test with nitrogen-doping technology, and normal operation gradient below 20 MV/m with the lower limit of $Q_0$ of $1.5 \times 10^{10}$ for long term operation), robust and variable high power (> 300 kW CW) input couplers that are design compatible with cavity clean assembly and low heat load, efficient and economical damping of the HOM power with minimum dynamic cryogenic heat load, and fast RF ramp and control of the Booster.

In parallel with design and key R&D, extensive development of SCRF personnel, infrastructure and industrialization is essential for the successful realization of CEPC. This will have synergy with ongoing SCRF-based accelerator projects in China, such as SHINE (Shanghai HIgh repetition rate XFEL aNd Extreme light facility) in Shanghai and ADANES (Accelerator Driven Advanced Nuclear Energy System) in Huizhou, etc.

**Collective effects**

Interaction of an intense charged particle beam with the vacuum chamber can lead to collective instabilities. These instabilities will induce beam quality degradation or beam losses, and finally restrict the performance of the machine.

The impedance and wake are calculated both with formulas as well as simulations with ABCI and CST. Both longitudinal and transverse impedances are dominated by the resistive wall and elements of which there is a large quantity. The longitudinal loss factor is mainly contributed by the resistive wall and the SCRF cavities.

The impedance driven instabilities include single bunch and multi-bunch effects. The limitation on the longitudinal broadband impedance mainly comes from microwave instability and bunch lengthening. The microwave instability will rarely induce beam losses, but may reduce the luminosity due to the distorted beam distribution and increasing of the beam energy spread. The limitation on the transverse broadband impedance mainly comes from the transverse mode coupling instability. Considering the bunch lengthening due to the longitudinal impedance, the transverse effective impedance will decrease with bunch intensity. The narrowband impedances are mainly contributed by cavity like structures. These impedances may induce coupled bunch instabilities in both longitudinal and transverse planes. The dominant contributions to the coupled bunch instability include resistive wall impedance and the HOM of the SCRF cavities. In order to damp the instabilities, bunch by bunch feedback systems are required. For different operation scenarios, the Z-pole mode has the most critical restrictions for both broadband and narrowband impedances.

Except the impedance driven effects, there are also two stream instabilities that may restrict the beam performance in the collider. In the electron ring, instabilities can be excited by residual gas ions accumulated in the potential well of the electron beam. Fast beam ion instability is a transient beam instability excited by the beam-generated ions accumulated in a single passage of the bunch train. The instability can be faster than the radiation damping. For the most critical case of Z-pole energy, multi-bunch train filling patterns and transverse feedback system are required.

Electron Cloud Effect is one of the considerations in positron ring of CEPC collider. The photon electrons and secondary electron emission will be the main contribution to the electron cloud. The most effective mitigation on electron cloud is coating TiN or NEG to depress secondary electron emission yield (SEY) to approximation one. The simulation for electron cloud density in different SEY and its threshold have been carried out during the conceptual design of CEPC. The shortest bunch spacing during Z-pole operation is 25 ns with bunch population $8.0 \times 10^{10}$. A transverse feedback system for damping ECI is also taken as a backup.

**CEPC booster**

The booster provides electron and positron beams to the collider at different energies. Both the initial injection from zero current and the top-up injection should be fulfilled. The booster is in the same tunnel as the collider, placed above the collider ring except in the interaction region where there are bypasses to avoid the detectors at IP1 and IP3.

The injection system consists of a 10 GeV Linac, followed by a full-energy Booster ring. Electron and positron beams are generated and accelerated to 10 GeV in the Linac, are injected into the Booster. The beams are then accelerated to full-energy, and injected into the Collider. For different beam energies of Higgs, W, and Z experiments, there will be different particle bunch structures in the Collider. To maximize the integrated luminosity, the injection system will operate mostly in top-up mode, but also has the ability to fill the Collider from empty to full charge in a reasonable length of time. A traditional off-axis injection scheme is chosen as a baseline design of the beam injection to the main collider, and a swap-out injection is given as another choice for Higgs injection.

The design goal for the booster optics is to make sure the geometry is the same as the collider and satisfy the requirements of beam dynamics. The total number of magnets and sextupole families is minimized taking into account capital and operating costs. The maximum cell length and hence the maximum emittance in the booster is limited by the collider injection requirements. The length of two FODO cells in the booster corresponds to three FODO cells in the collider. The horizontal position of the booster has been designed in the center of collider two beams. The height difference between booster and collider is 2.4m and the horizontal position error of booster is controlled under ± 0.17m. CEPC booster has the same circumference as the collider (100016.4m). 90°/90° FODO cell and non-interleave sextupole scheme is adopted to achieve the biggest Dynamic aperture (DA).

DA reduction due to sawtooth effect at 120GeV is negligible so magnets energy tapering is unnecessary in booster. The error analysis has been done and so far the errors in the booster are tolerable.

During ramping, parasitic sextupole field is induced on beam pipe inside dipoles due to eddy current. So the ramping rate (0.1Hz) is limited by eddy current effect. Dedicated ramping curve is found to control the maximum K2. Even if the dynamic chromaticity distortion is corrected by the independent sextupoles during ramping, the DA reduction around 20GeV is serious. A beam simulation of the entire period in the booster from injection to extraction is needed to verify the design, using a realistic beam from the linac.

Low field dipole magnet is aslo a big challenge for CEPC because the booster dipoles have to start from 29 Gauss. The requirement for field error field reproducibility at low energy is difficult to reach. Both technical solutions and physical solutions will be explored.

**CEPC linac injector**

The CEPC linac injector to booster is a normal conducting S-band linac with frequency 2860 MHz providing electron and positron beams at an energy of up to 10 GeV at a repetition rate of 100 Hz. One-bunch-per-pulse is adopted and bunch charge should be larger than 1.5nC.

In the design of CEPC linac, the reliability and availability of the linac injector was emphasized because it is one of the indispensable facilities. The S-band linac has a robust design based on well proven technologies, and a 15% overhead of accelerating structure and klystron is foreseen to provide margins. The linear type layout of CEPC linac is adopted and one electron transport line at an energy of 4 GeV is desiged to bypass the positron source and part accelerating section. All the lattice design and multi-particle simulations are conducted and the linac can meet all the requirements with errors.

A thermionic gun is adopted and can provide 11 nC electron beam for positron production. To keep the potential to meet higher requirements and possibility of updates in the future, the linac can provide bunch charge larger than 3 nC electron beam and posirton beam. The positron source is a conventional design with a tungsten target of 15 mm in length and adiabatic matching device of 6 T in peak magnetic field. The energy of electron beam for positron production is 4 GeV and rms beam size is 0.5 mm.

A 1.1 GeV damping ring with 58.8m circumference is adopted to reduce the transverse emittance of positron beam to suitably small value. Longitudinal bunch length control has provided to minimize wake field effects in the linac by a bunch compressor system after the damping ring. Both parameters for damping ring and bunch compressor have designed, also the lattice design is finished and the DA can fulfill the requirement.

**CEPC collider magnets**

There are 2466 dipoles, 3052 quadrupoles, 948 sextupoles and 2904 correctors in the CEPC collider ring. Theses conventional magnets occupy over 80% of the 100 km circumference, therefore, the cost and power consumption are two of the most important issues for magnet design. The 2384 dipoles and 2392 quadrupoles are designed to be dual aperture magnets to provide magnetic field for both beams separated by 350 mm. Several special technologies are used to reduce the cost of the magnets, including core steel dilution for dipoles and aluminum coils instead of copper. In addition, the magnets are designed for low-current high-voltage operating mode as much as possible to reduce the power consumption in the power cables. Radiation shielding is also considered. In the Technical Design Report (TDR) R&D phase, the dual aperture dipole and quadrupole short prototype magnets will be developed.

**Booster magnets**

The circumferences of the booster is about 100 km, which has 16320 dipoles, 2036 quadrupoles, 448 sextupoles and 350 correctors. The gap of the dipole magnets is 63 mm, the most of them are 4.7 m long, the others are 2.4 m and 1.7 m long. The field will change from 29 Gauss to 392 Gauss during acceleration. The field errors in good field region are required to be less than 1E-3. Due to very low field level, the cores are composed of stacks of 1 mm thick low carbon steel laminations spaced by 1 mm thick aluminium laminations. Since magnetic force on the poles is very small, the return yoke of the core can be made as thin as possible. In the pole areas of the laminations, some holes will be stamped to further reduce the weight of the cores as well as to increase the field in the laminations. All above considerations can improve the performance of the iron core and considerably reduces the weight and the cost. Also for economic reasons, the excitation bars are made from pure aluminum of cross section $30 \times 40$ mm$^2$ without water cooling.

The bore diameter of the quadrupole magnets is 64 mm, the magnetic length are 1 m, 1.5 m and 2.2 m respectively, the max. quadrupole field is 16.6 T/m. The min. quadrupole field is 1/12 of the max. field. For cost reduction, hollow aluminum instead of copper conductors are selected to wind the coils. The iron cores are made of 0.5 mm thick laminated low carbon silicon steel sheets. The magnet will be assembled from four identical quadrants, and can also be split into two halves for installation of the vacuum chamber.

The bore diameter of the sextupole magnets is also 64 mm, all the magnets are 400 mm long, the max. sextupole field is 440 T/m$^2$. The min. sextupole field is 1/12 of the max. field. The cores of the magnets have a two-in-one structure, made of low-carbon silicon steel sheets and end plates. By using end chamfering, the field errors can be reduced to meet the strict field requirements. The coils of the magnets have a simple racetrack-shaped structure, which are wound from solid copper conductors without water cooling.

The gap of the correctors is 63 mm, the max. field is 200 Gs, the field errors in good field region is required to be less than 1E-3. To meet the field quality requirements, the correctors have H-type structure cores so the pole surfaces can be shimmed to optimize the field. The cores are stacked from 0.5 mm thick laminations. The racetrack shaped coils are wound from solid copper conductor. Each coil has 24 turns, which are formed from 4 layers; no water cooling is required.

### CEPC interaction region superconducting magnets

Compact high gradient final focus superconducting quadrupole doublet magnet (QD0 and QF1) are required on both sides of the collision points of CEPC collider ring. QD0 and QF1 are double aperture quadrupoles and are operated fully inside the field of the Detector solenoid with a central field of 3.0 T. To minimize the effect of the longitudinal detector solenoid field on the accelerator beam, anti-solenoids before QD0, outside QD0 and QF1 are needed. The total integral longitudinal field generated by the detector solenoid and anti-solenoid coils is zero. It is also required that the total solenoid field inside the QD0 and QF1 magnet should be close to zero. The superconducting QD0, QF1, and anti-solenoid coils are in the same cryostat, which makes up a combined function magnet. In the TDR R&D phase, superconducting prototype magnets for the CEPC interaction region will be developed in three consecutive steps: 1) Double aperture superconducting quadrupole prototype magnet QD0 (136T/m, 2m long). 2) Short combined function superconducting prototype magnet including QD0 and the anti-solenoid. 3) Long combined function superconducting prototype magnet including QD0, QF1 (110T/m, 1.48 long) and anti-solenoid.

### CEPC vacuum system

R & D of the CEPC vacuum system include the vacuum chambers, RF shielding Bellows and NEG coating inside the inner surface of the vacuum system in the positron ring.

The Collider will have an aluminum chamber for the electron beam and a copper chamber for the positron beam. The aluminum chamber is similar to the LEP chamber. It has a beam channel, a cooling water channel, a pumping port used to install ion pump, and thick lead shielding blocks covering the outside. The copper chamber has a beam channel and a cooling water channel, and NEG coating will be used. In the R&D program both types of chambers will be fabricated and tested. The prototypes of both copper and aluminum vacuum chambers will be fabricated and tested. The ultimate pressure of the vacuum chambers is less than $2\times10^{-10}$ Torr；The thermal outgassing rate is less than $5\times10^{-12}$ Torr•L/s/cm$^2$.

The NEG coating is a titanium, zirconium, vanadium alloy, deposited on the inner surface of the chamber through sputtering. R&D is required so the sputtering process for NEG film deposition is optimized to avoid instability and lack of reproducibility. These problems can significantly change the gas sorption and surface properties (e.g. secondary electron yield, ion-induced gas desorption). During tests of the coating, all related parameters (plasma gas pressure, substrate temperature, plasma current, and magnetic field value) will be recorded and suitably adjusted to ensure stability of the deposition process. After coating, the chambers will be cooled down to room temperature, exposed to air and left to age for a couple of days before being visually inspected again. Aging is a recommended procedure, since it helps identify areas where the film adhesion is poor. The design specification with pumping speed of 0.5 L/s/cm$^2$ ($H_2$) for the coating vacuum chamber will be achieved.

The primary function of the RF shielding bellows is to allow for thermal expansion of the chambers and for lateral, longitudinal and angular offsets due to tolerances and alignment, while providing a uniform chamber cross section to reduce the impedance seen by the beam. The usual RF-shield has many narrow Be-Cu fingers that slide along the inside of the beam passage as the bellows is compressed. The design specification of the contact pressure for RF contact fingers is 125±25g/finger.

## CEPC 650MHz high efficiency klystron

The CEPC two beam synchrotron radiation power is more than 60 MW, it needs high efficiency RF source to minimize CEPC AC power consumption. Considering the klystron operation lifetime and power redundancy, a single 650MHz 800 kW klystron amplifier will drive two of the collider ring SC cavities through a magic tee and two rated circulators and loads. The CEPC high efficiency 650MHz klystron design goals are to set the efficiency to be above 80% and successful industrialization.

## Civil and conventional facilities

Underground structures of CEPC consist of collider ring tunnel (L=99.67km), experiment halls (2 experiment halls for CEPC, and 2 future experiment halls for SPPC), CEPC linac injector is located on the surface. Surface structures within the collider ring area, such as auxiliary equipment structures, cooling towers, substations and ventilation systems, are located close to the shafts. The total area of surface structures is 140450m$^2$. The total CEPC tunnel civil construction period is 54 months, including 8 months for construction preparation, 43 months for construction of main structures and 3 months for completion. The total electrical load for physical experiments and general facilities is 270MW. It is proposed to use 220kV for the project, and to have two 220kV central substations (220kV/110kV/10kV) in the project area. The total heat load absorbed by the cooling water system is 213MW during CEPC operation for Higgs physics experiments. There are 16 pump stations at each point of the ring and an additional one for Linac. The air conditioning cold load of the collider ring tunnel is about 6MW. The tunnel is divided into 32 sections. Each section is considered to be independent for the ventilation and primary return air conditioning system. Fire prevention and exhaust systems, hydrant and fire extinguisher systems, and fire detection and fire alarm systems are combined with building fire prevention and evacuation, to minimize fire hazards.

**CEPC cost, site selection, management and construction**

The cost of CEPC contains mainly three parts, civil engineering, accelerator with two detectors. As cost repartition (without detectors, for example), civil engineering 40%; collider magnets 12%; booster magnets 6%, vacuum system 9%; RF power system 6%; SRF system 4%; mechanical system 5%; instrumentation 4%; cryogenic system 4%; linac injector 2.5%; and others.

As for CEPC site selection, the technical criteria are roughly quantified as follows: earthquake intensity less than seven on the Richter scale; earthquake acceleration less than 0.1 g; ground surface-vibration amplitude less than 20 nm at 1–100 Hz; granite bedrock around 50–100 m deep, and others. The site-selection process started in February 2015，preliminary studies of geological conditions for CEPC's potential site locations have been carried out in Qinhuangdao and Xiongan in Hebei province; Shenshan special district in Guangdong province; Huangling county in Shanxi province; Huzhou in Zhejiang province; and Chuangchun in Jilin province, and all these sites satisfies the CEPC construction requirements.

The scheduled timeline for CEPC construction is that after the TDR finished in 2022, CEPC will start the civil construction. According to Chinese civil construction companies involved in the siting process, a 100 km tunnel will take less than five years to dig using drill-and-blast methods, which is followed by accelerator and detectors' installation. Around 2030, CEPC construction will be finished followed by machine operation.

The CEPC by nature is Chinese initiated international large science project, and the project will be participated, contributed and managed in an international way in all its level and in all of its process from CDR, TDR, construction and operation for physics experiments.

**SppC**

SppC (Super proton-proton Collider) as an integral part of the CEPC-SppC project, aims for energy-frontier discoveries, which will be a long-term development or more than twenty years from now after CEPC as an electron-positron collider will accomplish its main physics goals. However, it is important to include the SppC in the general CEPC study, as we do not wish that the CEPC design would not hinder the future development to add the SppC collider in the same tunnel. Other physics possibilities such as e-p collision that involves both CEPC and SppC beams should not be forgot during the general project planning.

As SppC is a long-term project with many technical challenges, it is necessary to pursue studies to work on critical physics and technological key issues. One of the key issues is high-field superconducting magnets of at least 12 T. The SppC baseline design of is to use 12 T high-temperature iron-based superconductors for high field dipole magnets to allow proton–proton collisions at a centre of mass energy of 75 TeV and a luminosity of $10^{35}$ cm$^{-2}$ s$^{-1}$.

About the accelerator physics study on SppC, the studies include: machine layout and lattice design with compatibility to CEPC; collimation method and machine protection scheme to handle the extremely high stored energy in beams; potential in collision luminosity and critical interaction regions, etc. As SppC collider needs an injector chain that consists of a series of powerful accelerators, current design of injector chain consists of four accelerators in cascade.

As for the timeline of SppC, till 2035 is CDR and R&D period; till 2040 is the Technical Design Report period, from 2040-2045 is SppC construction period, and SppC will be put to operation started from 2045.

**References:**
[1] CEPC-SppC Pre-CDR, http://cepc.ihep.ac.cn/preCDR/volume.html, 2015.
[2] CEPC-SppC Progress Report, http://cepc.ihep.ac.cn/Progress%20Report.pdf, 2017.
[3] The CEPC Conceptual Design Report, Vol I: Accelerator, arXiv: 1809.00285, http://cepc.ihep.ac.cn/CDR_v6_201808.pdf, 2018.

**Addendum :**

### The Planning of CEPC

The Chinese government has established a program to host China initiated international large science projects. A call for proposals is expected in early 2019. After a selection process, the Ministry of Science and Technology will cultivate 3-5 seed projects by 2020, from which 1-2 projects will be approved for construction. China will continue to identify and support further large science projects in the future. This program provides a natural path for the realization of the CEPC.

The current priority of the CEPC project is to secure its position as one of the seed projects of this program. The official approval of the CEPC project, at the earliest, could happen in 2022. The accelerator construction could start soon after the approval. At the same time, a call for detector Letters of Intent is planned. Two detectors will be selected and the International collaborations will be formed accordingly. The collaborations should deliver their Detector Technical Design Reports within two years after the starting of the construction of the accelerator.

In November 2018, the CEPC Conceptual Design Report was released and the CEPC International Advisory Committee provided its recommendations. According to which, the accelerator study group should complete its Technical Design Report by 2022. Meanwhile, collaborative R&D work on each of sub-detector systems will be the focus. To achieve this goal, two International Committees will be established. The first is an Accelerator Review Committee that advises on all matters related to the accelerator design and R&D including the Machine-Detector Interface and upgrade capabilities. The second is a Detector R&D committee that reviews and endorses the Detector R&D proposals from the international community, such that the international participants could apply for funds from their funding agencies and make effective and sustained contributions.

International collaboration is vital for the CEPC project. Active collaborations have been established between the domestic CEPC study groups and multiple international research institutions. Through these collaborations, many key challenges of the CEPC detector design and physics studies have been identified and being addressed through dedicated R&D programs.

A new organization structure is proposed to promote and accommodate the future international participation, see Figure 1. This structure reflects the discussion at the 2018 CEPC workshop and takes into account the recommendations of the International Advisory Committee. It is intended for the period from 2019 till the construction.

In this structure, all the building blocks will integrate the international participation. The Institution Committee writes the bylaws and makes major decisions on organizational issues. The national representatives interface with the National Funding Agencies and the corresponding institutions are represented in the Institution Committee. Supported by the Accelerator Review Committee and the Detector R&D Committee, the Project director is responsible for the coordination of studies at each group.

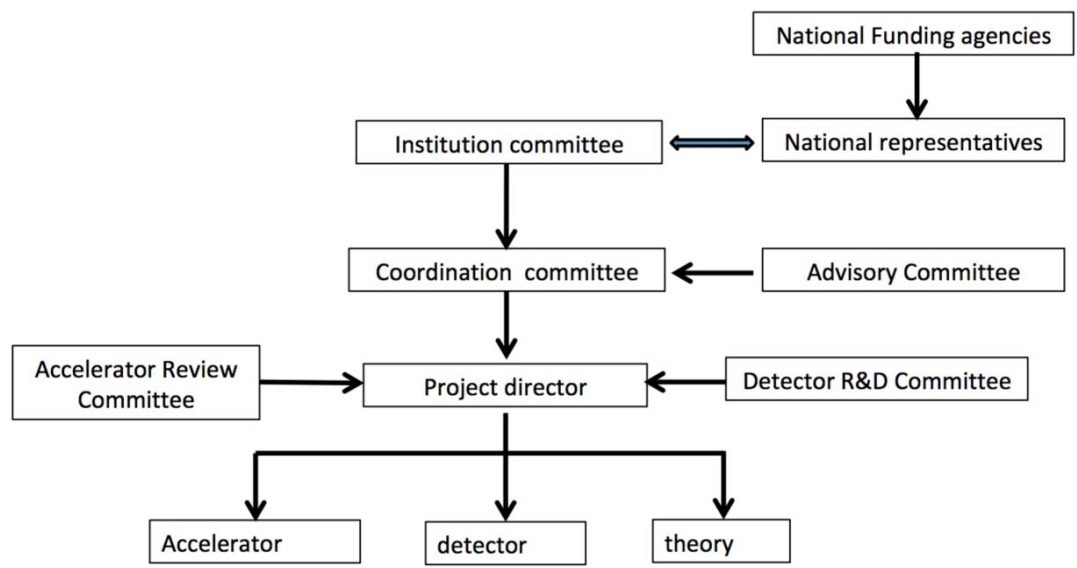

Figure 1: The planned international organization from 2019 till the construction

The organization will evolve with time. In the construction and operational phase, the organization will naturally evolve to include the council representing the participating countries and the host lab management who provide supervision to this project.

Though the CEPC project is initiated by and to be hosted in China, it is envisioned to be an international project. The organization and the management of the project will reflect the international participation of its stakeholders. The successful international participation played a critical role in the delivery of the CEPC conceptual design, and it will certainly become more important in the future.